\documentclass[letter,longauth,traditabstract]{aa}
\usepackage{txfonts}
\usepackage{graphicx}
\usepackage{natbib}
\bibpunct{(}{)}{;}{a}{}{,} 

\title{Herschel PEP: The star-formation rates of 1.5$<$z$<$2.5 massive galaxies
	\thanks{ Herschel is an ESA space observatory with science instruments provided by European-led Principal
Investigator consortia and with important participation from NASA.}
	}

\author{
R. Nordon \inst{\ref{mpe}}
\and D. Lutz \inst{\ref{mpe}} 
\and L. Shao \inst{\ref{mpe}}
\and B. Magnelli \inst{\ref{mpe}}
\and S. Berta \inst{\ref{mpe}}
\and B. Altieri\inst{\ref{ESA}}
\and P. Andreani\inst{\ref{ESO}, \ref{INAFtri}}
\and H. Aussel\inst{\ref{CEA}}
\and A. Bongiovanni\inst{\ref{Canarias},\ref{Laguna}}
\and A. Cava\inst{\ref{Canarias},\ref{Laguna}}
\and J. Cepa\inst{\ref{Canarias},\ref{Laguna}}
\and A. Cimatti\inst{\ref{Bologna}}
\and E. Daddi\inst{\ref{CEA}}
\and H. Dominguez\inst{\ref{INAFbol}}
\and D. Elbaz\inst{\ref{CEA}}
\and N.M. F\"orster Schreiber \inst{\ref{mpe}}
\and R. Genzel \inst{\ref{mpe}}
\and A. Grazian \inst{\ref{INAFroma}}
\and G. Magdis\inst{\ref{CEA}}
\and R. Maiolino\inst{\ref{INAFroma}}
\and A.M. P{\'e}rez Garc{\'i}a\inst{\ref{Canarias},\ref{Laguna}}
\and A. Poglitsch \inst{\ref{mpe}}
\and P. Popesso\inst{\ref{mpe}}
\and F. Pozzi\inst{\ref{INAFroma}}
\and L. Riguccini\inst{\ref{CEA}}
\and G. Rodighiero \inst{\ref{Padova}}
\and A. Saintonge\inst{\ref{mpe}}
\and M. Sanchez-Portal\inst{\ref{ESA}}
\and P. Santini\inst{\ref{INAFroma}}
\and E. Sturm\inst{\ref{mpe}}
\and L. Tacconi\inst{\ref{mpe}}
\and I. Valtchanov\inst{\ref{ESA}}
\and M. Wetzstein \inst{\ref{mpe}}
\and E. Wieprecht \inst{\ref{mpe}}
}

\institute{MPI for Extraterrestrische Physik, Postfach 1312, 85741 Garching, Germany \label{mpe}
\and Herschel Science Centre \label{ESA}
\and ESO, Karl-Schwarzschild-Str. 2, D-85748 Garching, Germany \label{ESO}
\and Department of Astronomy, University of Padova, Vicolo dell'Osservatorio 3, I-35122 Padova \label{Padova}
\and Instituto de Astrof{\'i}sica de Canarias, 38205 La Laguna, Spain \label{Canarias}
\and Departamento de Astrof{\'i}sica, Universidad de La Laguna, Spain \label{Laguna}
\and Laboratoire AIM, CEA/DSM-CNRS-Universit{\'e} Paris Diderot, IRFU/Service d'Astrophysique,
B\^at.709, CEA-Saclay, 91191 Gif-sur-Yvette Cedex, France \label{CEA}
\and Dipartimento di Astronomia, Universit{\`a} di Bologna, Via Ranzani 1,
40127 Bologna, Italy \label{Bologna}
\and INAF-Osservatorio Astronomico di Bologna, via Ranzani 1, I-40127 Bologna, Italy \label{INAFbol}
\and INAF - Osservatorio Astronomico di Roma, via di Frascati 33, 00040 Monte Porzio Catone, Italy \label{INAFroma}
\and INAF - Osservatorio Astronomico di Trieste, via Tiepolo 11, 34143 Trieste, Italy \label{INAFtri}
}

\date{Received <date> /Accepted <date>}

\abstract{
The star formation rate (SFR) is a key parameter in the study of galaxy evolution.
{The accuracy of SFR measurements at z$\sim$2 has been questioned following a disagreement between observations and theoretical models.}
The latter predict SFRs at this redshift that are typically a factor 4 or more lower than the measurements.
We present star-formation rates based on calorimetric measurements of the far-infrared (FIR) luminosities for massive 1.5$<$z$<$2.5, normal star-forming galaxies (SFGs), which do not depend on extinction corrections and/or extrapolations of spectral energy distributions.
The measurements are based on observations in GOODS-N with the Photodetector Array Camera \& Spectrometer (PACS) onboard {\it Herschel}, as part of the PACS Evolutionary Probe (PEP) project, that resolve for the first time individual SFGs at these redshifts at FIR wavelengths.
We compare FIR-based SFRs to the more commonly used 24~$\mu$m and UV SFRs.
We find that SFRs from 24~$\mu$m alone are higher by a factor of $\sim$4--7.5 than the true SFRs. This overestimation depends on luminosity: gradually increasing for $log$~L(24~$\mu$m)$>12.2$~L$_\odot$. The SFGs and AGNs tend to exhibit the same 24$\mu$m excess.
The UV SFRs are in closer agreement with the FIR-based SFRs. Using a Calzetti UV extinction correction results in a  mean excess of up to 0.3~dex and a scatter of 0.35~dex from the FIR SFRs.
The previous UV SFRs are thus confirmed and the mean excess, while narrowing the gap, is insufficient to explain the discrepancy between the observed SFRs and simulation predictions.
}


\titlerunning{SFR in 1.5$<$z$<$2.5 galaxies}
\authorrunning{Nordon et al.}

\begin{document}
\maketitle

\section{Introduction}
The evolution of massive galaxies around z$\sim$2, when the cosmic star-formation rate density was at its peak, is currently the subject of intense study. A key parameter in constraining their nature and stellar-mass buildup is the instantaneous star-formation rate (SFR). 
While there are various methods for estimating SFRs \citep[see ][ for a review]{Kennicutt98}, it is notoriously difficult to constrain SFRs of z$\sim$2 galaxies {with an accuracy better than a factor of a few}.
One of the reasons is that calibrations between observed luminosities (rest-frame UV, mid-IR, and submm being the most accessible and widely used) and bolometric SFRs are based on local galaxies and their validity at high z has not yet been firmly established.

Both UV (BX/BM) and optical/near-IR (BzK) selected samples of star-forming galaxies at z$\sim$2
have been successful in producing large samples for galaxy evolution studies \citep{Reddy05, Perez-Gonzalez05, Daddi07a, Santini09}. However, UV-based SFRs are very sensitive to dust obscuration and there are uncertainties about the applicable reddening laws and dust/source geometry at high~z \citep[e.g., ][]{Reddy10}.
Estimates from the 24~$\mu$m luminosities rely on template-based extrapolations to the total IR luminosity that are poorly constrained at high~z and may be affected by an AGN. Likewise, submm luminosities rely on assumptions about IR SEDs, which depend on poorly known dust temperatures, composition, and spatial distribution.

Cosmological simulations predict a tight correlation between stellar mass and SFR with a slope similar to that observed \citep[e.g.,][]{Finlator06}, but with SFRs that are a factor $\sim$4 lower than those inferred for z$\sim$2 massive galaxies from a combination of UV and 24~$\mu$m luminosities \citep{Daddi07a, Wilkins08, Damen09}. 
{Comparisons between the derivative of the mass density distribution and direct measurements of SFR density also lead to a discrepancy, in particular at z$>$2 \citep[e.g. ][]{Perez-Gonzalez08}.}
Suggestions on how to resolve this discrepancy include changes to the adopted initial mass functions \citep[IMF, ][]{Dave08} and star formation efficiency \citep{Khochfar09}, but these cannot be properly tested given the uncertainties associated with the common SFR indicators themselves.

The photo-detector array camera and spectrometer \citep[PACS, ][]{Poglitsch10} onboard the {\it Herschel} space observatory \citep{Pilbratt10} offers a new opportunity to infer the SFR of z$\sim$2 galaxies from their far infrared (FIR) emission. PACS observes at 160~$\mu$m (45--65~$\mu$m rest frame for the sample used here), close to the peak of the emission from dust heated by young stars, away from AGN-heated dust emission and with unprecedented spatial resolution that reduces confusion noise. It allows the most reliable and least biased SFR measurements to date, which avoid large SED extrapolation errors and avoid the uncertainties of attenuation corrections. 
In this letter, we report on the first measurements of FIR-based calorimetric SFRs of optically selected galaxies at 1.5$<$z$<$2.5 and compare them to results from previous UV and 24~$\mu$m measurements. 
We adopt a $\Omega_m = 0.3$, $\Omega_\Lambda = 0.7$, and $H_0 = 70$~km~s$^{-1}$~Mpc$^{-1}$ cosmology throughout this letter.

\section{Data, sample and SFR(160$\mu$m)}
\label{sec:data}
The PACS data consists entirely of PACS evolutionary probe (PEP) guaranteed-time observations in the GOODS-N field, taken during the science demonstration phase of the {\it Herschel} mission. For details of the observations and data reduction process, we refer to \citet{Berta10}. Fluxes were extracted using {\it Spitzer}-MIPS 24~$\mu$m sources with 3~$\sigma$ detections or better as priors \citep{Magnelli09}. 
{At the current PACS depth (5.7 mJy, 3$\sigma$, at 160~$\mu$m), testing showed that practically all PACS sources are detected at 24~$\mu$m. Hence, using 24~$\mu$m sources as priors allows a complete PACS source extraction.}
The priors are matched to bviz (ACS), JHK (FLAMINGO), and {\it Spitzer}-IRAC photometry. Spectroscopic redshifts are from \citet{Barger08} and photometric redshifts were derived using the code EAZY \citep{Brammer08}.
The {\it Chandra}~2Ms catalog \citep{Alexander03} is matched to the PACS sources and is used to flag X-ray sources.
{Using the \citet{Ranalli03} L$_x$/L$_{FIR}$ relation for SFG, the expected X-ray flux of a log(LIR/L$_\odot$)=12 galaxy at z=2 is almost an order of magnitude lower than the $1.4 \times 10^{-16}$~ergs~cm$^{-2}$~s$^{-1}$ hard-band (observed 2--8~keV) sensitivity limit of the 2~Ms catalog. We therefore treat every X-ray source with a hard-band detection as an AGN.}
One source without an X-ray detection but which exhibits a clear power-law SED in the IRAC bands, was also flagged as an AGN.

In this study, we are interested in 1.5$<$z$<$2.5 range and require redshift information. {Hence, only sources for which we have spectroscopic redshifts, and/or were successfully fitted by EAZY for a photometric redshift are considered in the following selections}. In addition, we require that in a 10" radius around a 24~$\mu$m prior there is no more than one other 24~$\mu$m source and it may have no more than 50\% of the prior flux. This condition ensures that the sample does not experience significant flux confusion between MIR priors and/or FIR fluxes due to neighbors. 
These criteria result in a sample of 23 detected sources in total, 17 classified as SFGs, and 4 of which have a spectroscopic redshift. 306 24~$\mu$m priors without a PACS detection are available for stacking analysis (see below).

We calculate SFRs by assuming that they are proportional to the integrated infrared luminosity \citep[LIR, typically 8-1000~$\mu$m][]{Kennicutt98}.
We convert PACS 160~$\mu$m fluxes into LIR by fitting the flux using the \citet[][hereafter: CE01]{CE01} SED library, in which for a given redshift and band flux, a unique solution to LIR exists.
The 100~$\mu$m filter is not used because at z$\sim$2, emission at rest-frame wavelengths may already be affected by hot dust and suffer the same systematic effects as the 24~$\mu$m based SFR. 
In addition, at these redshifts 100~$\mu$m is on the same side of the SED peak and too close to 160~$\mu$m to provide meaningful temperature constraints, without longer wavelength data. Combined PACS and SPIRE sub-mm data show that fitting CE01 SEDs to the 160~$\mu$m flux provides a reliable estimate of the total 8--1000~$\mu$m luminosity (LIR) of the large-grain dust for z$\sim$2 galaxies \citep{Elbaz10}.

To explore the relationship between direct FIR luminosities, 24~$\mu$m extrapolated luminosities, and UV-SFRs to lower luminosities and SFRs, we also perform a stacking analysis.
We stack the 160~$\mu$m image centered on the prior positions and measure a mean flux for the stack.
The mean luminosity of the stack, LIR(160~$\mu$m)$_{stack}$, is defined as the luminosity that 
will result in the 160~$\mu$m stacked flux, for the redshift distribution of the sources in the stack and the adopted CE01 SEDs.
In this way, both k-corrections and luminosity distances are accounted for without bias. The error in the mean luminosity is calculated using a bootstrap method. The same number of sources are resampled with replacements from the original stack to produce a new stacked image. The new mean luminosity is then calculated. The error is finally calculated from the distribution of luminosities in the repeated resampling. 

\section{24$\mu m$ based SFR}
\label{sec:24um}
\begin{figure}
\begin{center}
\includegraphics[width=0.9\columnwidth]{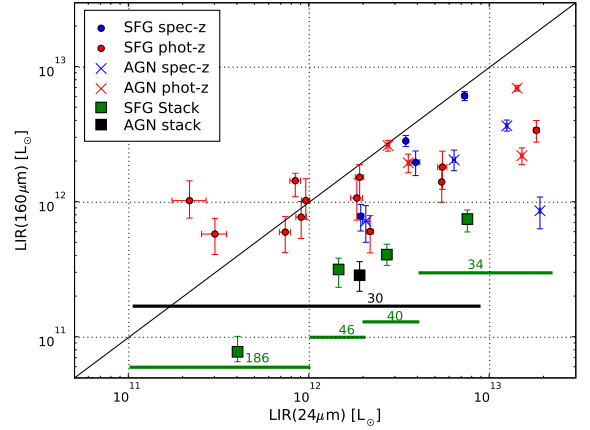} 
\caption{\label{fg:L24vsL160}
	Total (8-1000$\AA$) IR luminosity from 160~$\mu$m versus that derived from 24~$\mu$m. Detections with spec-z are in blue, detections with phot-z only are in red. SFGs are plotted as circles and AGNs as x marks. The error bars include only photometric errors. {Squares represent mean luminosity of stacked} SFG (green) and X-ray (black) undetected sources, with error bars indicating the error in the mean luminosity. Horizontal bars beneath the stacks indicate the min--max values in the stack with the number of stacked sources above them.}
\end{center}
\end{figure}

\begin{figure}
\begin{center}
\includegraphics[width=0.9\columnwidth]{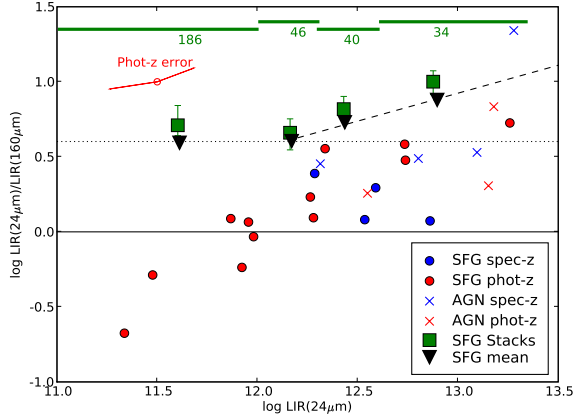} 
\caption{\label{fg:L24overL160}
	The log ratio of 24~$\mu$m to 160~$\mu$m total IR luminosities (proportional to SFR) as a function of LIR from 24~$\mu$m. 
	Colors and symbols are similar to Fig.~\ref{fg:L24vsL160} with black triangles representing the mean of all (detections and non-detections) SFGs in the stacked range. The red arrow at top left indicates the typical uncertainties caused by phot-z errors.}
\end{center}
\end{figure}

Spitzer-MIPS 24~$\mu$m fluxes are converted to LIR by fitting the 24~$\mu$m flux to CE01 SEDs using the same method we use for the PACS~160~$\mu$m fluxes (Sect.~\ref{sec:data}).
{This SED library is very commonly used in 24~$\mu$m (and other) studies where LIR is often derived from a single photometric point. Many other SED libraries exist, but the $\nu$L$_\nu$(8~$\mu$m)/LIR ratio that is relevant to this study is typically calibrated to match local galaxy observations and vary by factors smaller than those discussed below.}
At z$\sim$2, the rest-frame wavelengths are shorter than 10~$\mu$m and 24~$\mu$m fluxes probe the IR emission at the edge of the relevant range, far from the SED peak, making them sensitive to extrapolation errors. In addition, at these wavelengths polycyclic aromatic hydrocarbon (PAH) emissions contribute significantly and the ratio of their fluxes to LIR may create significant scatter. PACS 160~$\mu$m fluxes probe the emission much closer to the SED peak and are unaffected by the above issues.

In Fig.~\ref{fg:L24vsL160}, we plot LIR as derived from 160~$\mu$m versus that derived from 24~$\mu$m for 1.5$<$z$<$2.5. Individual sources are detected starting at LIR(160~$\mu$m)$>6\times 10^{11}$~L$_{\odot}$ at redshifts close to z$\sim$1.5. The 24~$\mu$m sources without a PACS detection (and no X-ray counterpart) are stacked in luminosity bins as indicated by the bars in the figure. {Stacked luminosities represent the mean luminosity as defined in Sect.~\ref{sec:data} and the stack error bars indicate the error in the {\it mean} LIR(160~$\mu$m) of the stack.}
Since we stack in luminosity and not flux bins, the typical redshifts of sources in a given stack are higher than those of detections with similar LIR(160~$\mu$m), which is one reason why not all sources in the luminosity bin are detected.
It is clear that while there is some scatter, for most sources LIR(24~$\mu$m) is higher than LIR(160~$\mu$m).

Figure~\ref{fg:L24overL160} plots the LIR(24~$\mu$m) to LIR(160~$\mu$m) ratio as a function of LIR(24~$\mu$m). The detections appear to show a linear trend, but this is mostly due to a selection effect. The combined mean for the detection and non-detections (detected in stacks) is much higher and the individual detections represent a fraction of the distribution for which at a given LIR(24~$\mu$m), LIR(160~$\mu$m) is high enough to be detected.
There is a weak 24~$\mu$m luminosity dependence and the mean of the excess increases with LIR(24~$\mu$m), at least for $\log$~LIR(24~$\mu$m)$>12.2$. Below this luminosity, it is harder to determine the slope, but the stack indicates that it may flatten out. 
{The dashed and dotted lines in Fig.~\ref{fg:L24overL160} are ad hoc fits to the binned combined-means slope above and a constant offset below LIR(24)=12.2}
\begin{equation}
\begin{array}{lcr}
0.38 \log \rm{LIR}(24) -4.0 & ; &\log \rm{LIR}(24)>12.2 \\
0.6 & ; &\log \rm{LIR}(24) \le 12.2 \\
\end{array}
\label{eq:L24correction}
\end{equation}

Though a large fraction of the sample relies on photometric redshifts, errors in redshifts will move the points in LIR(24$\mu$m), but will have a much smaller effect on the LIR ratio. The red arrows at the top-left of Fig.~\ref{fg:L24overL160} demonstrate how the typical phot-z errors, propagated into the quantities plotted in the figure, will affect the values. The photo-z induced dispersion is directed along the diagonal trend of detected sources and probably contributes to this pattern with the selection effects.

X-ray AGNs do not appear to differ from the rest of the sample in Figs.~\ref{fg:L24vsL160} and \ref{fg:L24overL160}, except perhaps for the one source with the highest luminosity. This source also exhibits a very clear power-law SED in the mid-IR. For sources completely dominated by the AGN, fitting a CE01 SED to the 24~$\mu$m emission is clearly inappropriate. 
Even though we can only produce one AGN stack that results in a good detection (Fig.~\ref{fg:L24vsL160}), the mean luminosity is similar to that of the SFG stacks. This can be interpreted in two ways: either the flux at $6.8<\lambda<9.6$~$\mu$m rest frame is dominated by the starburst emission at all the luminosities considered here, or the flux excess in all sources is due to an AGN, which is obscured in most.
The latter claim and the relatively tight LIR(24~$\mu$m)/LIR(160~$\mu$m) ratio would imply a tight relation between the AGN luminosity and the galaxy's starburst component FIR luminosity, which is not observed \citep{Lutz10, Shao10}. We conclude that while a hidden AGN may contribute to the scatter, it is not likely to be the main cause of the general 24~$\mu$m excess. Enhanced emission (relative to local galaxy SEDs) in PAH features between 6.2--8.6~$\mu$m, that enter the 24~$\mu$m filter at z=1.5 is a more plausible explanation.

These new results corroborate the main findings of \citet{Murphy09}, based on {\it Spitzer} 16, 24, 70 $\mu$m, and SCUBA 850~$\mu$m observations. For z$\sim$2 galaxies, they find that SFRs using only 24~$\mu$m are overestimated by a factor of 5 and conclude that a hidden AGN can only account for a fraction of this excess. 
\citet{Papovich07}, using {\it Spitzer} 24, 70 and 160 $\mu$m stacked fluxes also find that LIR(24~$\mu$m) is in excess. At the high end, they find 24~$\mu$m overestimates LIR by a factor of 2--10. They too conclude that X-ray sources have a 24~$\mu$m excess similar to SFGs without X-ray counterparts.

\section{UV-based SFR}

\begin{figure}
\begin{center}
\includegraphics[width=0.9\columnwidth]{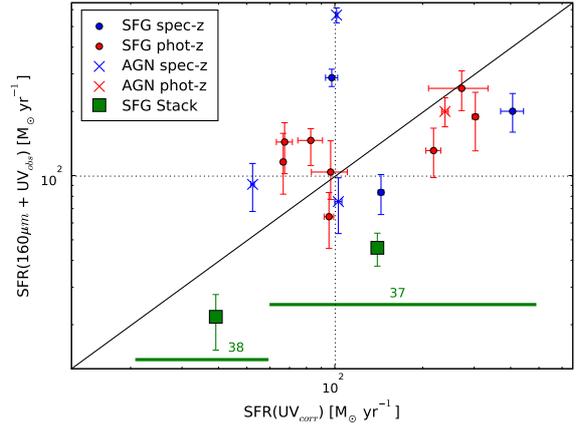} 
\caption{\label{fg:BzKvs160}
	The sum of SFR from 160~$\mu$m and UV$_{obs}$ (no attenuation correction) versus SFR derived from attenuation-corrected UV flux. Blue symbols are for spec-z, red are for phot-z. Circles are SFG detections, x-marks are AGNs. Errors in detected sources are photometric only. {Squares are mean SFR for stacks} with the error in the mean. Horizontal bars below indicate the min--max range of values in the stack with the number of stacked sources noted above.
	}
\end{center}
\end{figure}

\begin{figure}
\begin{center}
\includegraphics[width=0.9\columnwidth]{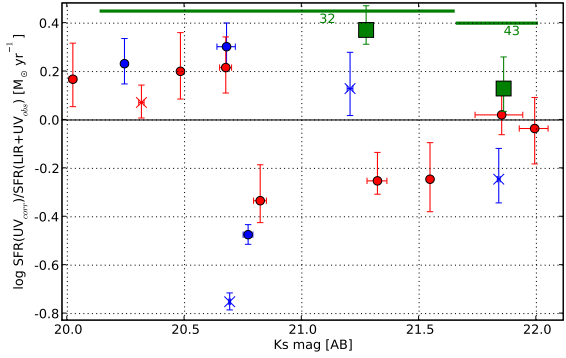} 
\caption{\label{fg:UVoverLIRvsKs}
	The log ratio of UV SFR to the combined LIR and UV$_{obs}$ SFR versus Ks magnitude. Colors and symbols are the same as in Fig.~\ref{fg:BzKvs160}.}
\end{center}
\end{figure}

The UV continuum at wavelengths longer than the Lyman edge can be used as a SFR indicator. The UV emission is dominated by young, massive, and short-lived stars and a SFR that is required to sustain their population can be calculated. 
The SFR(UV) is sensitive to the attenuation correction used, which varies significantly in the UV between different extinction curves, all derived from local galaxies. Testing the accuracy of SFR(UV) is essentially a test of the attenuation correction.
Here we use B and z band photometry (probing rest frame $\sim$1500~$\AA$ and $\sim$3500~$\AA$ at z$\sim$2) to estimate their SFR and dust attenuation.
For the optical extinction, we apply the calibration given by \citet{Daddi04}
\begin{equation}
E(B-V) = 0.25(B-z+0.1)_{\rm{AB}}.
\label{eq:EBV}
\end{equation}
The effective attenuation at 1500~$\AA$ is obtained using the \citet{Calzetti00} extinction law $A_{1500}=10E(B-V)$, which was derived for local starburst galaxies.

The BzK sample is selected according to the \citet{Daddi04} color criterion. We require at least 3~$\sigma$ detection in all 3 bands, a cut at Ks$<$22~AB, and a redshift (spectroscopic or photometric) 1.5$<$z$<$2.5. PACS 160~$\mu$m fluxes were extracted using 24~$\mu$m priors, which implies a 24~$\mu$m detection for this selection.
It also means that the BzK sample is a subsample of that used in Sect.~\ref{sec:24um}.
The sample includes 15 sources detected at 160~$\mu$m, 11 classified as SFG. 
Seventy-five sources without 160~$\mu$m detections are used for stacking. Strictly speaking, the requirement for a 24~$\mu$m prior is irrelevant to both SFR(UV) and SFR(160) and may even bias the selection against sources with lower obscuration. To improve the stacked signal, we also include BzK sources without a 24~$\mu$m detection, but verify that introducing this requirement gives consistent results.
In converting from luminosities to SFRs, we follow the \citet{Kennicutt98} relations, but lower the SFR by a factor 1.7 to convert from a Salpeter 
initial mass function (IMF) to that of \citet{Chabrier03}, though for the results discussed here the choice of IMF makes little difference:
\begin{eqnarray}
SFR_{LIR} [M_\odot / yr^{-1}] = \frac{LIR}{9.85\times 10^{9}\, \rm{L_\odot}}; \\
SFR_{UV} [M_\odot / yr^{-1}] = \frac{L_\nu(1500\AA)}{1.5 \times 10^{28}\, \rm{erg\, s^{-1} Hz^{-1}}}.
\label{eq:SFR}
\end{eqnarray}
In Fig.~\ref{fg:BzKvs160}, we plot the combined SFR from LIR(160~$\mu$m) and the observed (uncorrected) UV flux versus the attenuation-corrected UV SFR, SFR(UV$_{corr}$). The LIR is assumed to be caused by the reprocessing of radiation from young stars by the dust and the addition of the SFR from the observed UV luminosity accounts for the escaped part of the UV radiation. In our sample, the latter makes only a small contribution with a mean of 2.4\% and maximum of 8\% of the total SFR for PACS sources.
Our detection limit with PACS is at about 70~M$_\odot$~yr$^{-1}$, below which we can only study sources by means of stacking. However, for a cut level $\rm{Ks}<22$ the small number of BzK galaxies below the PACS threshold limits the stacking possibilities. The detections exhibit good agreement between SFR(UV$_{corr}$) and the combined SFR(LIR+UV$_{obs}$). 
The mean log-ratio of the detections is -0.02~dex and standard deviation of 0.25~dex.
We stack PACS non-detections by SFR(UV$_{corr}$) bins. For the low SFR stack, the two SFR estimators are in very good agreement, while for the other, the value of SFR(UV$_{corr}$) is slightly too high. The overall mean log-ratio of stacks and detections is 0.3~dex and the standard deviation is 0.35~dex.
{We therefore conclude that when using the Calzetti UV extinction law for SFR(UV)$\gtrsim$40~M$_\odot$~yr$^{-1}$  the SFR is overestimated by a mean factor $\sim$2 with a scatter of a similar magnitude.}

The scatter in Fig.~\ref{fg:BzKvs160} exhibits a systematic pattern. For SFR(UV$_{corr}$)$>$100~M$_\odot$~yr${-1}$, this estimator tends to overpredict the SFR of {the PACS {\it detections}} by a factor$\sim$1.5 and below this rate to underpredict by roughly the same factor. This trend is also seen in Fig~\ref{fg:UVoverLIRvsKs}, where the SFR indicator ratio is plotted vs. the Ks magnitude. 
Since the pattern appears to be also a function of Ks magnitude (and of both B and z mag to a lesser degree), it is unlikely to be caused by phot-z errors. It is not preserved when plotting against either redshift or extinction. 
{Both stacks for the Ks magnitude bins lie above the zero-line, indicating a spread in the full population and shift the overall mean upwards.} 
The difference between stacks and detections is likely a selection effect: for a given SFR(UV) (or Ks), one will generally tend to detect in 160~$\mu$m galaxies for which the attenuation is underpredicted. However, the clear dependency on both SFR and Ks magnitude and the observed trend for the stacks might suggest there is more to it.
The geometry and metallicity of the galaxies may play a role in this. \citet{Reddy10} suggested that some young galaxies may have a different UV reddening law, although these galaxies represent only 13\% of their sample. The current sample is too small to draw conclusions about these effects.

Four X-ray AGNs that are selected by BzK were detected with PACS at 160~$\mu$m. Three are indistinguishable from the SFG in Figs.~\ref{fg:BzKvs160}~\&~\ref{fg:UVoverLIRvsKs} and both their UV and FIR emission must be dominated by the star formation. The fourth (log-ratio=-0.75) is an extremely bright AGN and both SFR indicators are probably not applicable to this case. This is the only 160~$\mu$m source that does not appear in Figs.~\ref{fg:L24vsL160} and \ref{fg:L24overL160} as it is too bright in 24~$\mu$m to be fit by CE01 SEDs. There were not enough non-detected AGN-BzKs to produce a meaningful stack. The presence of an AGN, except in the extreme cases, does not significantly affect the SFR(UV).

\bibliographystyle{aa} 
\bibliography{14621bibli}
\begin{acknowledgements}
PACS has been developed by a consortium of institutes led by MPE (Germany) and including UVIE
(Austria); KU Leuven, CSL, IMEC (Belgium); CEA, LAM (France); MPIA (Germany); INAF-IFSI/
OAA/OAP/OAT, LENS, SISSA (Italy); IAC (Spain). This development has been supported by the
funding agencies BMVIT (Austria), ESA-PRODEX (Belgium), CEA/CNES (France), DLR (Germany),
ASI/INAF (Italy), and CICYT/MCYT (Spain).
\end{acknowledgements}

\end{document}